\documentclass[aps,prl,twocolumn,superscriptaddress,showpacs]{revtex4}
\usepackage{graphicx}

\usepackage{dcolumn}

\usepackage{bm}
\usepackage{color}
\usepackage[colorlinks]{hyperref}
\input{epsf}

\begin{document}


\title{Observation of a uniform temperature dependence in the electrical resistance across the structural phase transition in thin film vanadium oxide ($VO_{2}$)}

\author{R. G. Mani}
\email{mani.rg@gmail.com} \affiliation{Georgia State University,
Department of Physics and Astronomy, Atlanta, GA 30303}
\affiliation{Harvard University, School of Engineering and Applied
Sciences, Cambridge, MA 02138}
\author{S. Ramanathan}
\affiliation{Harvard University, School of Engineering and Applied
Sciences, Cambridge, MA 02138}

%
%
%
\date{\today}
\begin{abstract}
An electrical study of thin  $VO_{2}$ films in the vicinity of the
structural phase transition at $68^{0}C$ shows (a) that the
electrical resistance $R$ follows $log (R)$ $\propto$ $-T$ over
the $T$-range, $20 < T < 80 ^{0}C$ covering both sides of the
structural transition, and (b) a history dependent hysteresis loop
in $R$ upon thermal cycling. These features are attributed here to
transport through a granular network.

\end{abstract}
%
%
\maketitle The $VO_{2}$ phase of the transition-metal-oxide
vanadium-oxide undergoes a structural monoclinic- to tetragonal-
phase transition in the vicinity of $68 ^{0}C$, which brings with
it a band-realignment and an insulator-to-metal transition including, in high quality single crystals,
orders-of-magnitude changes in the electrical resistance, and a
dramatic alteration in the infrared optical
transmission.\cite{1,2,3,4,5,6} This interesting
electronic-phase-transition
has been attributed both to the Peierls-type period doubling
lattice distortion, and an electron interaction induced Mott
transition.\cite{2,3,4,5} From a practical perspective, these
large changes in the electrical and optical properties near room temperature have suggested $VO_{2}$ as a smart material in future electro-optic applications. Studies have already shown that a large applied electrical
bias\cite{7,8}  or intense photoexcitation\cite{9,10,11,12} can
trigger the electrical transition even below the structural
transition temperature. Indeed, the electrical transition can be very
fast, $\approx 10^{2} fs$, in thin films and small
particles.\cite{10,11,12} This
confluence of practical and scientific interest have motivated
investigations that aim to better understand the basics, and simultaneously obtain control over the insulator-metal transition, independent of the
structural transition.

Here, we characterize and study thin films, which have thus far
generally exhibited less dramatic changes than single crystals
over the above mentioned transition(s). Thus, thin films of
$VO_{2}$ were deposited by electron beam evaporation of $VO_{2}$
pellets, onto photolithographically patterned rectangular windows
ranging from 200$\mu m$ x 400 $\mu m$ down to 50$\mu m$ x 12.5
$\mu m$, on R-plane sapphire substrates.\cite{13} The film
thickness was 350 nm. After $VO_{2}$ liftoff, Cr/Au was thermally
evaporated onto contact-pad-windows in the photoresist, and this
was followed by metal liftoff. The contact fabrication was
completed by a slow thermal anneal at $450^{0} C$. The reported
$R$ vs. $T$ and the $I-V$ measurements were carried in a
two-terminal configuration. Typically, the resistance was
collected during both the up sweep and the down sweep of $T$.
$I-V$ measurements were carried out concurrently at a series of
fixed $T$, with the bias ramped in both directions; these results
were insensitive to the bias sweep direction.
\begin{figure}[t]
\begin{center}
\leavevmode \epsfxsize=3in
 \epsfbox {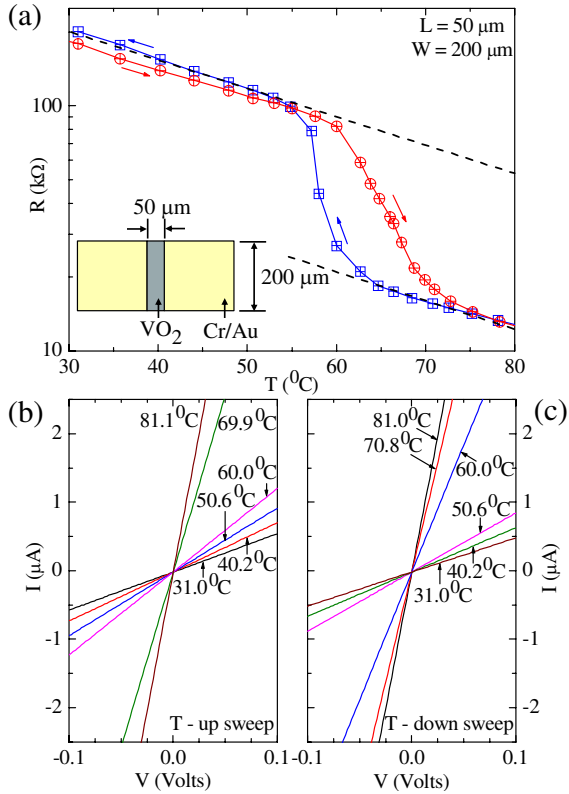}
\end{center}
\caption{Figure 1)(a): The two-terminal resistance, $R$, of a $50
\mu m$ long $\times$ $200 \mu m$ wide  $\times$ $350 nm$ thick
$VO_{2}$ device is shown versus the temperature, $T$. The drop in
$R$ in the vicinity of $60 < T < 75 ^{0}C$ with increasing $T$ is
a signature of the temperature-induced structural-transition in
this material. Note the hysteresis loop formed by the up-sweep and
down-sweep traces in the figure. The dashed lines convey the
approximate $log(R) \propto -T$ relation on both sides of the
structural transition in these specimens. The inset exhibits the
device geometry. (b) and (c): $I - V $ curves are shown for a
series of temperatures for the up sweep of the temperature, (b),
and the down sweep of the temperature, (c).} \label{mani01fig}
\end{figure}
Figure 1(a) shows the resistance, $R$,  of a $50\mu m$ long
$\times$ $200\mu m$ wide film. This trace shows the electrical
switching from a monoclinic,  high resistance, nominally
"insulating" state to a tetragonal, lower resistance, nominally
"metallic" state, and the switching occurs occurs at a higher
temperature on the up sweep- than on the down sweep- of
$T$.\cite{1} Other notable features observed in Fig. 1(a) are (i)
a difference in the base temperature resistance at the outset- and
at the end- of the measurement, (ii) an approximate linear
decrease of the resistance with increasing $T$ on the low-$T$,
high resistance side of the transition in the $log (R)$ vs $T$
plot, and (iii) a similar $T$ dependence of $R$ on the high-$T$,
low-resistance side of the structural transition.

Concurrent $I-V$ measurements are shown in Fig. 1(b) and 1(c),
with Fig. 1(b) and Fig. 1(c) showing the low bias $I-V$ data
obtained during the $T$ up sweep and $T$ down sweep, respectively.
Over the measured range, the current $I$ increases linearly with
the applied bias and the extracted $R$ are consistent with the
Fig. 1(a). An observable feature in the measurements of Fig. 1(b)
and Fig. 1(c) is that the slope of the $I-V$ curves changes
rapidly with $T$ over the phase transition, and the associated
$T$-interval is dependent upon the $T$ sweep direction, consistent
with the hysteretic transport of Fig. 1(a).

 Devices of different lengths were photolithographically fabricated on the same substrate in order to facilitate length-dependence studies. Measurements are exhibited in Fig. 2(a), for a 400$\mu m$, a
200$\mu m$, and a $100 \mu m$ long strip, with a width, $W
= 200 \mu m$. The noteworthy features in these data are once again that (i) in
each case, the room temperature resistance after thermal cycling
exceeds the initial room temperature resistance at the outset of
the experiment, (ii) the width in $T$ of the hysteresis loop does
not depend upon the length of the device, see also Fig. 2(b),
(iii) in the low temperature phase, i.e., below $T = 60^{0}C$, the
$T$ variation of $R$ is linear in this $log (R)$ vs $T$
plot, similar to the
data in Fig. 1(a), and (iv) the same behavior continues onto the
high temperature tetragonal, nominally "metallic"  $VO_{2}$ phase, immediately above the structural transition.

Identical devices on the same substrate also showed similar $log (R) \propto -T$ behavior on both sides of the structural transition; this is illustrated by the data of Fig. 3(a) for a pair of $200 \mu m$
long $\times$ $100 \mu m$ wide devices.

As mentioned above, these measurements indicate an increase of the
base temperature resistance upon thermal cycling. A close
examination of such behavior revealed, in addition, a change in
the width of the hysteresis loop, as exhibited in Fig. 3(b),
especially over the first few thermal cycles. Figure 3(b) shows
that the width of the hysteresis loop in the initial thermal cycle
formed by the up sweep trace $1$  and the down-sweep trace $2$
measurements is roughly a factor of two smaller than width of the
hysteresis loop observed in the second thermal cycle constituted
by traces $3$ and $4$. The figure also shows that $R(40^{0}C)$ in
these measurements increases as one proceeds from data trace $1$
to $2$ to $3$ to $4$. Thus, there appears to be a correlation
between the increase in the base temperature $R$ upon thermal
cycling and the broadening of the hysteresis loop over the range
of $T$ characterizing the structural transition.

\begin{figure}[t]
\begin{center}
\leavevmode \epsfxsize=3in
 \epsfbox {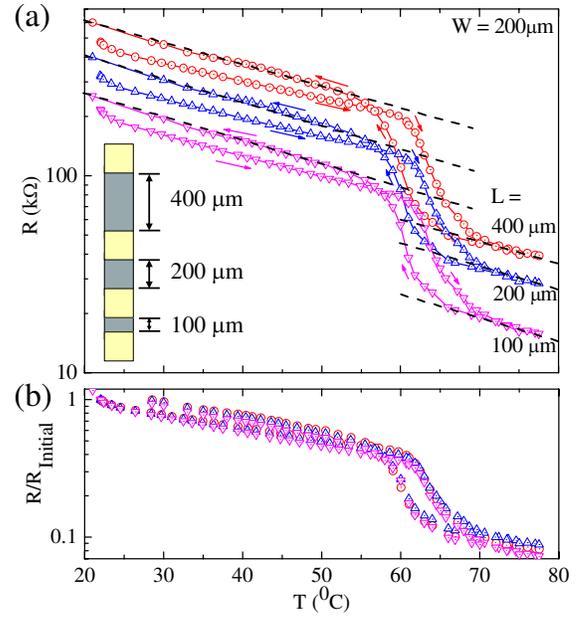}
\end{center}
\caption{Figure 2) (a): The two-terminal resistance, $R$, of $200
\mu m$ wide $\times 350 nm$ thick $VO_{2}$ devices is shown versus
the temperature, $T$, for several lengths of the device. The drop
in $R$ in all three devices between $60 < T \leq  70 ^{0}C$ with
increasing $T$ is a signature of the $T$-induced
structural-transition in this material. Note the similar
hysteresis loop in all three devices. The exhibited data traces
were collected simultaneously. The dashed lines convey the
approximate $log(R) \propto -T$ relation on both sides of the
structural transition. (b) The normalized resistance
$R/R_{Initial}$ is shown vs. $T$ for the $400 \mu m$, $200 \mu m$,
and the $100 \mu m$ long devices of (a).} \label{mani01fig}
\end{figure}

We present here a qualitative explanation, which builds upon earlier results. Berglund and Guggenheim reported
long ago that single crystals of $VO_{2}$, and many other
materials that exhibit a structural transition, include a
tendency to fissure upon passing through the phase
transition.\cite{4} The literature also demonstrates a strong
correlation between the width of the hysteresis loop and the size
of $VO_{2}$ particles or grains in the $VO_{2}$ system. Most
recently, it has been shown that ensembles of nanometer sized
particles exhibit a substantial hysteresis loop upon cycling
through the transition,\cite{14} and this behavior was attributed to the reduced
number of transition-triggering nucleation sites in
smaller particles, and the consequent spread in transition
temperatures in an ensemble of such small particles. The previous
report of a wider hysteresis loop in smaller particles suggests
the interpretation here that the increasing width of the
hysteretic loop with repeated thermal cycling, see Fig.
3(b), originates from a reduction in the average grain size in thin films upon thermal cycling. Thus, the electrical response of these $VO_{2}$ films seems to be a signature of transport through a granular network, and the increase in the base
temperature resistance with thermal cycling seems to reflect the
increased resistance in a grain-boundary dominated percolating
network, as the grains become smaller, within a given device.
\begin{figure}[t]
\begin{center}
\leavevmode \epsfxsize=2.8in
 \epsfbox {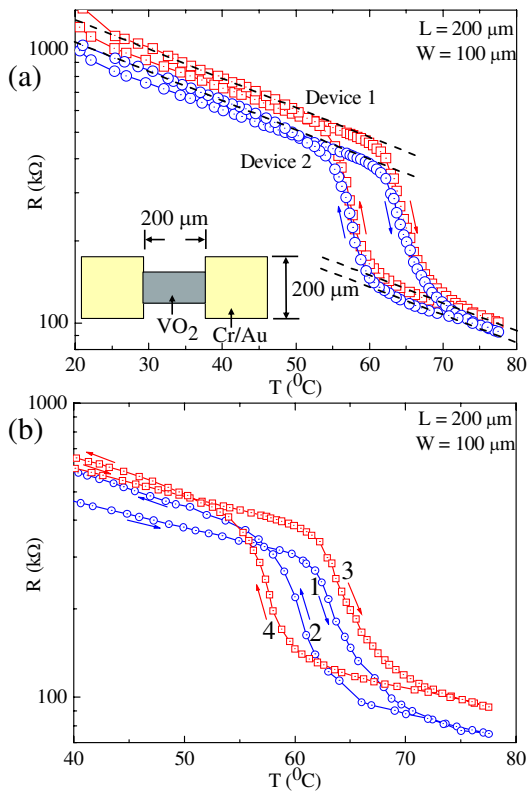}
\end{center}
\caption{Figure 3) (a) The two-terminal resistance, $R$, is shown
versus the temperature, $T$, for a pair of $200 \mu m$ long
$\times$ $100 \mu m$ wide $\times$ $350 nm$ thick $VO_{2}$
devices, on the same substrate. The measurements were carried out
simultaneously. (b) Two consecutive thermal cycles of $R$ vs. $T$
are shown for a single $200 \mu m$ long $\times$ $100 \mu m$ wide
$\times$ $350nm$ thick $VO_{2}$ device. Note the change in the
width of the hysteresis loop between the first thermal cycle which
is made up of the traces labelled $1$ and $2$, and the second
thermal cycle which is made up of the traces $3$ and $4$.}
\label{mani02fig}
\end{figure}
The other remarkable feature of these measurements on thin films
is the similarity of the $T$-dependence of $R$ on both sides of
the structural transition. This feature is in sharp contrast to
the behavior observed in high quality single crystals, where the
structural transition brings with it also a metal-insulator
transition.\cite{4} Note that the insulating state often exhibits
activated transport, $R \propto exp(\epsilon/k_{B}T)$, and a
divergent resistance in the $T \rightarrow 0K$ limit, while the
metallic state typically exhibits a finite $R$ in the
$T\rightarrow 0K$ limit. The observation in these experiments of a
similar non-activated \textit{and} non-metallic $T$-dependence on
\textit{both} sides of the structural transition is therefore
surprising. This feature suggests, perhaps, the absence of a true
metallic state in these thin films.

We suggest here the conjecture that effective localization through
the formation of small grains helps to transform the nominally
"metallic" state into a disorder mediated, relatively high
resistance state with the $log (R) \propto -T$ characteristics.
The hypothesis was advanced long ago that, in disordered single
crystals, "the so-called metallic phase [in $VO_{2}$] is not
metallic at all".\cite{3} In that instance, Mott\cite{3} was
referring to data\cite{1} from specimens showing just a one-to-two
orders-of-magnitude change in $R$ over the structural transition.
Perhaps, the disordered thin films examined in this study are
physically similar to the single crystals then considered. The
uniform temperature dependence  then appears plausible here,
however, if the $T$-dependence of the coupling within the granular
network sets the temperature dependence of the resistance on both
sides of the structural transition.

Work at Harvard was supported by NSEC-Harvard under NSF PHY-0601184.


\begin{thebibliography}{19}

\section{references}
\bibitem{1} F. Morin, Phys. Rev. Lett. \textbf{3}, 34 (1959).

\bibitem{2} D. Adler,  Rev. Mod. Phys. \textbf{40}, 714 (1968).

\bibitem{3} N. F. Mott,  Rev. Mod. Phys. \textbf{40}, 677 (1968).

\bibitem{4} C. N. Berglund and H. J. Guggenheim, Phys. Rev. \textbf{185}, 1022 (1969).

\bibitem{5} V. S. Vikhnin, S. Lysenko, A. Rua, F. Fernandez,
and H. Liu, Phys. Lett. A \textbf{343}, 446 (2005).

\bibitem{6} G. Golan, A. Axelevitch, B. Sigalov, and B.
Gorenstein, J. Optoelect. and Adv. Materials, \textbf{6}, 189
(2004).

\bibitem{7} P. P. Boriskov, A. A. Velichko, and G. B. Stefanovich, Phys. Sol. St. \textbf{46}, 922
(2004).

\bibitem{8} H-T Kim, B-G. Chae, D-H. Youn, G. Kim, and K-Y. Kang, Appl. Phys. Lett. \textbf{86},
242101 (2005).

\bibitem{9} M. F. Becker, A. B. Beckman, R. M. Walser, T. Lepine, P. Georges, A. Brun, Appl.
Phys. Lett. \textbf{65}, 1507 (1994).

\bibitem{10} R. Lopez, T. E. Haynes, L. A. Boatner, L. C. Feldman,
and R. F. Haglund, Jr., Opt Lettl. \textbf{27}, 1327 (2002).

\bibitem{11} A. Cavalleri, Th. Dekorsy, H. H. W. Chong, J. C. Kiefferr, and R. W. Schoenlein,
Phys. Rev. B \textbf{70}, 161102 (2004).

\bibitem{12} M. Rini, A. Cavalleri, R. W. Schoenlein, R. Lopez, L. C. Feldman, R. F. Hagland, L. A. Boatner, and T. E. Haynes, Opt. Lett., \textbf{30}, 558 (2005).

\bibitem{13} H. Bialas, A. Dillenz, H. Dowmar, and Pi. Ziemann, Thin Solid
Films \textbf{338}, 60 (1999).

\bibitem{14} R. Lopez, T.E. Haynes, L. A. Boatner, L. C.
Feldman, and R. F. Haglund, Jr., Phys. Rev. B \textbf{65}, 224113
(2002).

\end{thebibliography}
\end{document}